\documentstyle{aipproc}

\begin{document}
January, 1997 \hfill{IFT-P/009/97}\\
\phantom{bla}
\hfill{solv-int/9701013}

\title{The structures underlying soliton \\ solutions in  integrable
hierarchies\footnote{Talk given at the I Latin American Symposium on High
Energy Physics, I SILAFAE, Merida, Mexico, November/96}}

\author{Luiz A. Ferreira\footnote{Partially supported by a CNPq research
grant}}
\address{Instituto de F\'\i sica Te\'orica - IFT/UNESP\\
Rua Pamplona 145\\
01405-900, S\~ao Paulo - SP, Brazil}

\maketitle

\begin{abstract}

We point out that a common feature of integrable hierarchies presenting soliton
solutions is the existence of some special ``vacuum solutions'' such that the 
Lax
operators evaluated on them, lie in some abelian subalgebra of the associated 
Kac-Moody algebra. 
The soliton solutions are constructed out of those ``vacuum solitons'' by the
dressing transformation procedure. 

\end{abstract}

This talk is concerned with the structures responsible for the appearance of
soliton solutions for a large class of non linear differential equations.  
In spite of the great variety of types of  equations presenting soliton 
solutions, some basic features seem to be common to all of them. 
Pratically all such theories  have a representation in terms of a zero 
curvature condition \cite{ferreira:zc}, and the corresponding Lax operators lie 
in some
infinite dimensional Lie algebra, in general a Kac-Moody algebra 
${{\hat {\cal G}}}$. We argue that
one of the basic ingredients for the appearence of soliton solutions in such
theories is the existence of ``vacuum solutions'' 
corresponding to Lax operators lying in some abelian (up to central term)
subalgebra of ${{\hat {\cal G}}}$. Using the dressing transformation procedure
\cite{ferreira:dress} 
we construct
the solutions in the orbit of those vacuum solutions, and conjecture that the
soliton solutions  correspond to some  special points in
those orbits. The talk is based on results obtained in collaboration with J.L.
Miramontes and J. Sanchez Guill\'en and reported in ref. \cite{ferreira:1}.

We consider non-linear integrable hierarchies of equations which 
can be formulated in terms of a
system of first order differential  
\begin{equation}
{{\cal L}}_N \Psi = 0\, , \qquad 
{{\cal L}}_N  \equiv {\partial \, \over \partial t_N} - A_N 
\label{ferreira:LaxGen}
\end{equation}
where the variables $t_N$ are the various ``times'' of the hierarchies, 
and their
number may be finite or infinite. The equations of the hierarchies are then 
equivalent to the integrability or zero-curvature conditions 
of (\ref{ferreira:LaxGen})
\begin{equation}
[{\cal L}_N\>, \>{\cal L}_M]=0\>. 
\label{ferreira:ZeroCurv}
\end{equation}
Therefore, the Lax operators are ``flat connections''
\begin{equation}
A_N = {\partial \Psi\over \partial t_{N}}\>
\Psi^{-1}\>.
\label{ferreira:psidef}
\end{equation}

The type of integrable hierarchy considered here is based on a Kac-Moody
algebra ${{\hat {\cal G}}}$, furnished with an integral gradation 
\begin{equation}
{{\hat {\cal G}}} = 
\bigoplus_{i\in{{{\relax\leavevmode}\hbox{\sf Z\kern-.4em Z}}}} 
{{\hat {\cal G}}}_i\> \qquad 
{\rm and}\qquad 
[{{\hat {\cal G}}}_i,{{\hat {\cal G}}}_j] \subseteq 
{{\hat {\cal G}}}_{i+j}\>. 
\label{ferreira:Gradation}
\end{equation}
The connections are of the form 
\begin{equation}
A_N = \sum_{i=N_{-}}^{N_{+}} A_{N,i} \> , \quad  {\rm where}\quad
A_{N,i}  \in {{\hat {\cal G}}}_i
\label{ferreira:genpot}
\end{equation}
where $N_{-}$ and $N_{+}$ are non-positive and non-negative integers, 
respectively.

We assume that the hierarchy posssesses at least one vacuum solution such that
the connections evaluated on such solution has the form
\begin{equation}
A_N^{({\rm vac})}\> =\>  \sum_{i=N_{-}}^{N_{+}} c_N^i b_i+ f_N (t)\> C
\>\equiv  \> \varepsilon_N + f_N (t)\> C\>. 
\label{ferreira:vacpot}
\end{equation}
where $C$ is the central element of ${{\hat {\cal G}}}$ and $b_i \in 
{{\hat {\cal G}}}_i$ are generators of a subalgebra  of 
${{\hat {\cal G}}}$ satisfying
\begin{equation}
[b_j,b_k] = j \>\beta_j\>C\> \delta_{j+k,0}
\label{ferreira:oscila}
\end{equation}
with $\beta_j$ being some complex numbers ($ \beta_{-j} = \beta_j$),  
$c_N^i$ are constants, and  $f_N (t)$ are functions ot the times 
$t_N$.~\footnote{As a consequence of (\ref{ferreira:ZeroCurv}), those functions 
have to satisfy 
${\partial\> f_N(t)\over \partial\> t_M} \> -\> {\partial\> f_M(t)\over 
\partial\> t_N}\> = \>
\sum_{i}\> i\> \beta_i\> c_M^i\> c_N^{-i}\>$} 
These vacuum potentials correspond to the solution of the associated linear
problem (\ref{ferreira:LaxGen})-(\ref{ferreira:psidef}), given by the group  
element  
\begin{equation}
\Psi^{({\rm vac})} = \exp \left( \sum_{N} \varepsilon_N t_N
\> +\> \gamma (t) \> C \right)
\label{ferreira:vacelem}
\end{equation}
where the numeric function $\gamma (t)$ is a solution of the equations
\begin{equation}
{\partial \gamma (t) \over \partial t_N} = f_N (t) + {1\over 2}\sum_{M,i}
i\>\beta_i\> c_N^i\> c_M^{-i}\>t_M\>. 
\label{ferreira:gamma}
\end{equation}

We now consider the dressing tranformations which map known solutions of the
hierarchy into new solutions \cite{ferreira:dress}. Denote by 
${{\>\hat{G}\>}}_-$, 
 ${{\>\hat{G}\>}}_+$, and ${{\>\hat{G}\>}}_0$ the subgroups of the Kac-Moody 
group ${{\>\hat{G}\>}}$ formed
by exponentiating the subalgebras ${{\hat {\cal G}}}_{<0} \equiv 
\bigoplus_{i<0}
{{\hat {\cal G}}}_i$, ${{\hat {\cal G}}}_{>0}\equiv \bigoplus_{i>0} 
{{\hat {\cal G}}}_i$, and 
${{\hat {\cal G}}}_0$,
respectively. According to Wilson~\cite{ferreira:WILa}, the 
dressing
transformations can be described in the following way. Consider a solution
$\Psi$ of the linear problem~(\ref{ferreira:LaxGen}), and let 
$\rho= \rho_-\> \rho_0\> \rho_+$ be a constant element in the ``big cell'' of  
${{\>\hat{G}\>}}$, {\it i.e.\/}, in the subset 
${{\>\hat{G}\>}}_-\> {{\>\hat{G}\>}}_0\> {{\>\hat{G}\>}}_+$ of 
${{\>\hat{G}\>}}$, such   
that 
\begin{equation}
\Psi\> \rho\> \Psi^{-1} = (\Psi\> \rho\> \Psi^{-1})_{<0}\> (\Psi\> \rho\>
\Psi^{-1})_0\> (\Psi\> \rho\> \Psi^{-1})_{>0}\>.
\label{ferreira:Factor}
\end{equation}
Notice that these conditions are equivalent to say that both $\rho$ and $\Psi\> 
\rho\> \Psi^{-1}$ admit a generalized Gauss decomposition with respect to the
gradation (\ref{ferreira:Gradation}). Define 
\begin{eqnarray}
\Psi^{\rho} \> & = &\> {{\Theta_{-}^{(0)}}} [(\Psi\> \rho \> 
\Psi^{-1})_{<0}]^{-1}\> \Psi \rho \equiv \Theta_{-}\,\Psi \,\rho
\nonumber\\
&  = &\> {{\Theta_{+}^{(0)}}} \> (\Psi\> \rho\> \Psi^{-1})_{>0} \>\Psi\> 
\equiv \Theta_{+}\, \Psi  
\label{ferreira:Dressing}
\end{eqnarray}
where
\begin{equation}
{{{\Theta_{-}^{(0)}}}}^{-1}\, {{\Theta_{+}^{(0)}}}  = 
\left( \Psi \rho \Psi^{-1}\right)_{0}.
\label{ferreira:tp0tm0}
\end{equation}

Then, $\Psi^{\rho}$ is another solution of the linear problem. Indeed, one can
check that by exploring the gradation of ${{\hat {\cal G}}}$ and the fact that 
the dressing
transformation is written in two ways (\ref{ferreira:Dressing}), that the 
transformed
connection lie in the same subspace of ${{\hat {\cal G}}}$ as the original one,
i.e. 
\begin{equation}
A_{N}^{\rho} = {\partial \Psi^{\rho}\over \partial t_N}\> ({\Psi^{\rho}})^{-1}
\in \bigoplus_{i=N_-}^{N_+}{{\hat {\cal G}}}_{i}\>,
\label{ferreira:DressZero}
\end{equation}

We now consider the orbit of the vacuum solution (\ref{ferreira:vacelem}) 
under the group  
of dressing transformations. For each constant group element $\rho$ one gets a
new solution, out of the vacuum solution, by the transformation 
$\Psi^{(\rm vac)}\> \mapsto\> \Psi^{\rho} = 
\Theta_{-}\> \Psi^{(\rm vac)} \rho = 
\Theta_{+}\> \Psi^{(\rm vac)}$. The the vacuum connection $A_N^{({\rm vac})}$ 
becomes
\begin{eqnarray}
A_{N}^{\rho} - f_N(t)\> c \> &=&\> {\Theta_{-}}\> \varepsilon_N \> 
{{\Theta_{-}}}^{-1}\> +\> 
\partial_{N}{\Theta_{-}}\>{{\Theta_{-}}}^{-1}\> \in\> 
\bigoplus_{i\leq N_+}
{\hat {\cal G}}_{i}\nonumber\\ 
\> &=&\> {\Theta_{+}}\> \varepsilon_N \> {{\Theta_{+}}}^{-1}\> +
\> \partial_{N} 
{\Theta_{+}}\>{{\Theta_{+}}}^{-1} \>\in\> \bigoplus_{i\geq N_-} 
{\hat {\cal G}}_{i} ,  
\label{ferreira:DPlus}
\end{eqnarray}

The components $A_{N,i}$ in (\ref{ferreira:genpot}), of the connection 
are functionals of 
the fields of the hierarchy. Then one can consider (\ref{ferreira:DPlus}) 
as a local
change of variables, and use it to relate the
parameters of the group elements $\Theta_{+}$ and $\Theta_{-}$, to the fields
of the hierarchy. In fact, one can choose a suitable set of parameters to write
the fields in terms of them. The value of that particular set of parameters
evaluated on the solution can be obtained by considering matrix elements of the
form 
\begin{equation}
\langle{\mu}\mid \> {{\Theta_{-}}}^{-1}\> {\Theta_{+}}  \> 
\mid{\mu^{\prime}}\rangle = 
\langle{\mu}\mid\> e^{\sum_{N} \varepsilon_N t_N} \> \rho \> 
e^{-\sum_{N} \varepsilon_N t_N} \> \mid{\mu^{\prime}}\rangle\>,
\label{ferreira:solspec}
\end{equation}
where $\mid{\mu}\rangle$ and $\mid{\mu^{\prime}}\rangle$ are vectors in a given 
representation 
of ${\hat {\cal G}}$. The appropriate set of vectors is specified by 
the condition that  all
the required components of $\Theta_{-}$  and $\Theta_{+}$, used to parametrize
the fields, can  
be expressed in
terms of the resulting matrix elements. It turns out \cite{ferreira:1} that the 
required matrix elements, considered as functions of the
group element $\rho$, constitute the generalization of the Hirota's            
tau-functions for these hierarchies \cite{ferreira:HIR}.  
Moreover, Eq.~(\ref{ferreira:solspec}) is the analogue of the, so called,
solitonic specialization of the Leznov-Saveliev solution proposed in 
\cite{ferreira:TUROLA,ferreira:SOLSPEC,ferreira:fms,ferreira:SAVGERV} for
the  affine (abelian and non-abelian) Toda theories.

Consider now the common eigenvectors of the adjoint action of the
$\varepsilon_N$'s that specify the vacuum solution~(\ref{ferreira:vacpot}). 
Then, the
important class of multi-soliton solutions is conjectured to
correspond to group elements $\rho$ which are the product of
exponentials of eigenvectors
\begin{equation}
\rho = e^{F_1} \> e^{F_2} \> \ldots e^{F_n} \>, \qquad 
[ \varepsilon_N \> , \> F_k ] = \omega_N^{(k)} \> F_k \> , \quad k=1,2, 
\ldots n\>.
\label{ferreira:eigenb}
\end{equation}
In this case, the dependence of the solution upon the times $t_N$ can be made
quite explicit
\begin{equation}
\langle{\mu}\mid\> \Theta_{-}^{-1}\> \Theta_{+}  \> \mid{\mu^{\prime}} \rangle= 
\langle{\mu}\mid\> \prod_{k=1}^n \exp (e^{\sum_{N} \omega_{N}^{(k)} t_N} F_k )  
 \> \mid{\mu^{\prime}}\rangle\>.
\label{ferreira:solspecb}
\end{equation}
We emphasize that not all solutions of the type~(\ref{ferreira:solspecb}) 
are soliton
solutions, but we conjecture that the soliton and multi-soliton solutions are
among them.  The conjecture that multi-soliton solutions are
associated with group elements of the form~(\ref{ferreira:eigenb}) naturally 
follows from 
the
well known properties  of the multi-soliton solutions of affine Toda equations
and of hierarchies of the KdV type, and, in the sine-Gordon theory, it has been
explicitly checked in ref.~\cite{ferreira:BABT}. Actually, in all these cases, 
the
multi-soliton solutions are obtained in terms of representations of the 
``vertex 
operator'' type where the corresponding eigenvectors are nilpotent. Then, for
each eigenvector $F_k$ there exists a positive integer number $m_k$ such that
$(F_k)^{m} \not=0$ only if $m\leq m_k$. This remarkable property simplifies the
form of~(\ref{ferreira:solspecb} ) because it implies that
$e^{F_k}\> =\> 1\>+ \> F_k\> + \cdots+ (F_k)^{m_k}/m_k!$, which provides a
group-theoretical justification of Hirota's method \cite{ferreira:HIR}.   
 
An interesting feature of the dressing transformations method is the
possibility of relating the solutions of different integrable equations.
Consider two different integrable hierarchies whose vacuum solutions are
compatible, in the sense that the corresponding vacuum Lax operators commute.
Then, one can consider the original integrable equations as the restriction of
a larger hierarchy of equations. Consequently, the solutions obtained through
the group of dressing transformations can also be understood in terms of the
solutions of the larger hierarchy, which implies certain relations among them.
(see section~4 of \cite{ferreira:fms} for more details). 

The developments described here lead to a quite general definition of tau 
functions for 
such hierachies, in terms of integrable highest weight representations of the
associated Kac-Moody algebra \cite{ferreira:1}. 

\vspace{.5 cm}

\noindent{\bf Acknowledgements}

The author is very grateful to the organizers of I SILAFAE and VII EMPC, and 
to CLAF (Centro Latino Americano de F\'\i sica) for the finantial support.

\end{document}